\documentclass[aps,showpacs,prx,twocolumn,superscriptaddress]{revtex4}
\usepackage{epsfig,amsmath}
\usepackage{subfigure}
\usepackage{graphicx}
\usepackage{dcolumn}
\usepackage{stmaryrd}
\usepackage{mathrsfs}
\usepackage{pifont}
\usepackage{amsthm}
\usepackage{amssymb}
\usepackage{bm}
\usepackage{latexsym}
\usepackage{hyperref}

\newcommand{\beq}{\begin{equation}}
\newcommand{\eeq}{\end{equation}}
\newcommand{\beqa}{\begin{eqnarray}}
\newcommand{\eeqa}{\end{eqnarray}}

\usepackage{color}

\begin{document}

\title{Quantum correlations in bulk properties of solids obtained from neutron scattering}
\author{Ben-Qiong Liu}
\email{LosenQ@gmail.com}
\affiliation{Key Laboratory of Neutron Physics, Institute of Nuclear Physics and Chemistry, CAEP, Mianyang 621900, Sichuan Province, China}

\author{Lian-Ao Wu}
\email{lianaowu@gmail.com}
\affiliation{Department of Theoretical Physics and History of Science, The Basque Country
University (EHU/UPV), PO Box 644, 48080 Bilbao, Spain\\
IKERBASQUE, Basque Foundation for Science, 48011 Bilbao, Spain}

\author{Guo-Mo Zeng}
\affiliation{College of Physics, Jilin University, Changchun 130012,
People's Republic of China}

\author{Jian-Ming Song}
\affiliation{Key Laboratory of Neutron Physics, Institute of Nuclear Physics and Chemistry, CAEP, Mianyang 621900, Sichuan Province, China}

\author{Wei Luo}
\affiliation{Key Laboratory of Neutron Physics, Institute of Nuclear Physics and Chemistry, CAEP, Mianyang 621900, Sichuan Province, China}

\author{Yang Lei}
\affiliation{Key Laboratory of Neutron Physics, Institute of Nuclear Physics and Chemistry, CAEP, Mianyang 621900, Sichuan Province, China}

\author{Guang-Ai Sun}
\affiliation{Key Laboratory of Neutron Physics, Institute of Nuclear Physics and Chemistry, CAEP, Mianyang 621900, Sichuan Province, China}

\author{Bo Chen}
\affiliation{Key Laboratory of Neutron Physics, Institute of Nuclear Physics and Chemistry, CAEP, Mianyang 621900, Sichuan Province, China}

\author{Shu-Ming Peng}
\affiliation{Key Laboratory of Neutron Physics, Institute of Nuclear Physics and Chemistry, CAEP, Mianyang 621900, Sichuan Province, China}

\pacs{03.67.-a, 75.10.Jm, 78.70.Nx}

\begin{abstract}
We demonstrate that inelastic neutron scattering technique can be used to indirectly detect and measure the macroscopic quantum correlations quantified by both entanglement and discord in a quantum magnetic material, VODPO$_4\cdot\frac{1}{2}$D$_2$O. The amount of quantum correlations is obtained by analyzing the neutron scattering data of magnetic excitations in isolated V$^{4+}$ spin dimers. Our quantitative analysis shows that the critical temperature of this material can reach as high as $T_c=82.5$ K, where quantum entanglement drops to zero. Significantly, quantum discord can even survive at $T_c=300$ K and may be used in room temperature quantum devices. Taking into account the spin-orbit (SO) coupling, we also predict theoretically that entanglement can be significantly enhanced and the critical temperature $T_c$ increases with the strength of spin-orbit coupling.
\end{abstract}

\maketitle

\section{I. INTRODUCTION}

Quantum entanglement has received considerable attention in the field of quantum information science since it can be employed as a resource to perform tasks in quantum cryptography, quantum teleportation and even quantum computation \cite{R.Horodecki,S.M.Aldoshin,Z.-M.Wang}. Although entanglement is responsible for many quantum processes, it is fragile due to the inevitable interaction between a system and its environment, which makes it difficult to be used in quantum techniques \cite{T.Yu}. In the past few years, another measure of quantum correlations, quantum discord \cite{H.Ollivier}, has been studied extensively \cite{T.Werlang,P.Giorda,L.Mazzola,B.-Q.Liu,X.-X.Yi}, which quantifies nonclassical correlations in a quantum system including those not captured by entanglement. Non-zero quantum discord in separable mixed states can even be a resource for computational speedup and quantum enhancement \cite{X.-M.Lu}. In view of these potential technique applications, the detection and quantification of quantum correlations become vital in the experimental context. Up to now, various criteria and solutions to this problem have been proposed, such as Bell inequalities \cite{J.S.Bell,A.Aspect}, entanglement witnesses \cite{B.Terhal,M.Lewenstein}, measurement of nonlinear properties of a quantum state \cite{F.A.Bovino}. The experimental technique used in photon systems \cite{P.Kok} was thought to be promising. However, it has been found that the method is limited by the count rates, which makes it incapable of detecting some entangled states generated in the laboratory. Besides, environmental noise is also a significant disturbance in detecting weakly entangled states.

On the other hand, an oxavanadium phosphate compound, VOHPO$_4\cdot\frac{1}{2}$H$_2$O has attracted intense interests from the experimental chemical community, as well known to be a precursor of industrial catalysts for oxidizing \textit{n}-butane to maleic anhydride \cite{G.Centi,F.Trifiro,E.Bordes}. Besides, in solid state physics, its intriguing magnetic properties have been investigated with different experimental techniques \cite{D.A.Tennant,D.C.Johnston,Y.Furukawa,J.Kikuchi,J.W.Johnson}, including NMR, magnetic susceptibility measurements etc., and theoretically analyzed on the basis of \textit{ab initio} calculations \cite{M.Roca,H.J.Koo}. Each V$^{4+}$ exhibits a local spin-$\frac{1}{2}$. Several magnetic models have been proposed in the literatures, e.g., the isolated dimer model and the alternating chain model with debatable magnetic interaction scheme and spin exchange coupling strength \cite{S.Petit,S.Petit2004}.

The local spin-$\frac{1}{2}$ structure implies that this material can be a promising carrier of quantum bits in quantum information processing and its quantumness deserves an overall examination. In this paper, we study carefully the quantum correlations in the spin dimer material VODPO$_4\cdot\frac{1}{2}$D$_2$O, based on the experimental data of its magnetic excitations by using inelastic neutron scattering (INS) \cite{D.A.Tennant}. We show that quantum entanglement remains in this material at relatively high temperature $T_c=$82.5 K, which is much higher than that of the previously reported CN [Cu(NO$_3$)$_2$2.5D$_2$O] (about 5 K) \cite{C.Brukner,G.Xu,M.Yurishchev}. On the other hand, for quantum discord, the critical temperature can be even higher than 300 K, indicating that discord can be used as a resource for room temperature quantum devices.  It is clear that this peculiarity has a great practical implication for quantum information processing.

This paper is arranged as follows. Section II describes the magnetic model for VODPO$_4\cdot\frac{1}{2}$D$_2$O and briefly introduces the method to calculate quantum correlations. Section III is devoted to the relationship between quantum correlations and the inelastic neutron scattering experimental data. We also analyse the behavior of quantum correlations under the influences of temperature $T$ and the spin-orbit interaction, respectively. We conclude in Section IV.

\section{II. Model and quantum correlations}

The nearest-neighbor Heisenberg magnet can be modelled by \cite{J.T.Haraldsen}
\begin{equation}
H=\sum_{i}{J_{i} \mathbf{S}_i\cdot\mathbf{S}_{i+1}},
\end{equation}
where the sum runs over nearest-neighbor pairs, the exchange coupling constants $\{J_i\}$ are positive for antiferromagnetic and negative for ferromagnetic interactions, and $\mathbf{S}_i$ is the quantum spin operator for a spin-1/2 ion at site $i$.

The simplest spin cluster model is the spin dimer shown in Figure 1, which consists of isolated pair of magnetic ions described by the two-spin Heisenberg Hamiltonian $H=J\mathbf{S}_1\cdot\mathbf{S}_2$.  VOHPO$_4\cdot\frac{1}{2}$H$_2$O has been well described by this spin dimer model, as reported in Refs. \cite{J.W.Johnson,H.J.Koo,J.T.Haraldsen}, and supported by experiments \cite{D.A.Tennant}. Its density matrix in thermal equilibrium is given by \cite{J.-M.Cai,X.Wang}

\begin{displaymath}
\rho(T)=\frac{1}{4}
\begin{pmatrix}
1+G &  &   &  \\
  & 1-G & 2G  &  \\
  & 2G  & 1-G &  \\
  &   &    & 1+G
\end{pmatrix},
\end{displaymath}
with $G(T)=\frac{1-\exp{(J/k_BT)}}{3+\exp{(J/k_BT)}}$, and $G=\frac{4}{3}\langle\mathbf{S}_1\cdot\mathbf{S}_2\rangle$, where $\langle\mathbf{S}_1\cdot\mathbf{S}_2\rangle\equiv\langle S_1^x S_2^x\rangle+\langle S_1^yS_2^y\rangle+\langle S_1^zS_2^z\rangle$ is the sum of correlations of three orthogonal directions $x$, $y$, and $z$ \cite{E.Lieb}. The isotropy of Heisenberg interaction in spin space results in $\langle S_1^xS_2^x\rangle=\langle S_1^yS_2^y\rangle=\langle S_1^zS_2^z\rangle$. Here, the spin-spin correlation function $\langle\mathbf{S}_1\cdot\mathbf{S}_2\rangle$ can be employed as an entanglement witness, because for any product state of a pair of spins the correlation satisfies \cite{C.Brukner}
\begin{multline}
|\langle\mathbf{S}_1\cdot\mathbf{S}_2\rangle|=|\langle S_1^x\rangle\langle S_2^x\rangle+\langle S_1^y\rangle\langle S_2^y\rangle\\
+\langle S_1^z\rangle\langle S_2^z\rangle|\leq |\mathbf{S}_1||\mathbf{S}_2|\leq 1/4.
\end{multline}
This bound, acting as a quantum-classical boundary in solid-state physics, is thus an entangle-separable boundary. The upper bound is given by the Cauchy-Schwarz inequality $|\mathbf{S}|\equiv\sqrt{\langle S^x\rangle^2+\langle S^y\rangle^2+\langle S^z\rangle^2}\leq1/2$. Nonlocality is another aspect of quantum correlations, and can be indicated by a measure of violation of Bell inequalities, such as the Clauser-Horne-Shimony-Holt (CHSH) inequality $|\langle\mathbf{B}\rangle|\leq2$ \cite{J.F.Clauser},
\begin{equation}
\mathbf{B}=N_1N_2+N_1N'_2+N'_1N_2-N'_1N'_2,
\end{equation}
where $N_i$ and $N'_i$ ($i$=1,2) denote the operators corresponding to measurements on site $i$ \cite{G.Jaeger}. When $|\langle\mathbf{B}(\rho)\rangle|>2$, the state $\rho$ cannot be described by local realistic theories. For the density matrix $\rho(T)$, the maximal violation of Bell's inequality is given by $|\langle\mathbf{B}(\rho)\rangle|=2\sqrt{2}|G|$.

Now, we come to quantum entanglement and discord. The quantum mutual information between the two spins reads
\begin{multline}
\mathcal{I}(\rho)=\frac{1}{4}\Big[(1-3G)\log_2{(1-3G)}\\
        +3(1+G)\log_2{(1+G)}\Big],
\end{multline}
and the classical correlation is \cite{L.Henderson}
\begin{multline}
\mathcal{C}(\rho)=\frac{1}{2}\Big[(1-G)\log_2{(1-G)}\\
        +(1+G)\log_2{(1+G)}\Big].
\end{multline}
Therefore, it is straightforward to obtain the quantum discord as $\mathcal{D}(\rho)=\mathcal{I}(\rho)-\mathcal{C}(\rho)$. The pairwise entanglement can be measured by concurrence \cite{W.K.Wootters}, defined as
\begin{equation}
\mathbb{C}(\rho) = \textrm{\textrm{max}}({0, \sqrt{\lambda_1}-\sqrt{\lambda_2}-\sqrt{\lambda_3}-\sqrt{\lambda_4}}),
\end{equation}
where $\lambda_\tau$ ($\tau=1,2,3,4$) are the eigenvalues of the product matrix $R=\rho\tilde{\rho}$ in descending order with $\tilde{\rho}=(\sigma^y\otimes\sigma^y)\rho(\sigma^y\otimes\sigma^y)$.
In our case, the concurrence is
\begin{equation}
\mathbb{C}=\max{\{0, |G|-\textstyle{\frac{1}{2}}|1+G|\}}.
\end{equation}

\begin{figure}[tbp]
\includegraphics[width=8.0cm]{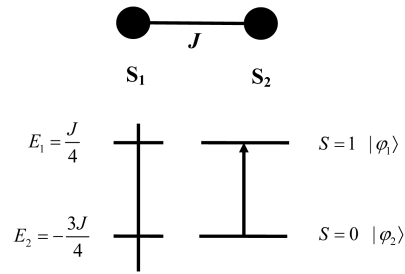}
\caption{The geometry and energy levels of a Heisenberg spin dimer. \label{Fig1}}
\end{figure}

\section{III. Quantum correlations obtained indirectly from INS experiment}

For magnetically ordered solids, the inelastic neutron scattering is particularly informative. In a magnetically coupled dimer, the total spin $S_{\rm{tot}}$ is a good quantum number and we can expect a dispersionless single excitation from the $S_{\rm{tot}}=0$ singlet to the excited level ($S_{\rm{tot}}=1$ triplet at $\Delta E=J$), which can be observed by inelastic neutron scattering. In such an experiment, the scattered neutrons are analyzed according to the energy transfer \cite{G.Shirane}
\begin{equation}
\hbar\omega=\frac {\hbar^2}{2m_n}\left(k_i^2-k_f^2 \right)
\end{equation}
where $m_n$ is the neutron mass, and $k_i$ and $k_f$  are the wave numbers of initial and final neutrons, respectively. The momentum transfer $\hbar\mathbf{Q}=\hbar(\mathbf{k}_i-\mathbf{k}_f)$ should be satisfied, where $\mathbf{Q}$ is the scattering vector. Assuming that both ground singlet and triplet states can be described by state vectors like $|S_1S_2SM\rangle$, where $S_1=S_2=1/2$ and $S$ is the total spin quantum number of the coupled system, the thermal neutron cross section of a transition $|S\rangle\rightarrow|S'\rangle$ can be written as \cite{H.U.Gudel}
\begin{multline}
\frac{d^2\sigma}{d\Omega d\omega}=\frac{N}{Z}\left( \frac{g\gamma r_0}{2}\right)^2\frac{k_f}{k_i}F^2(Q)e^{-2W}\cdot e^{-\frac{E_S}{k_BT}}\times\\
                                  \sum_{\alpha,\beta}{\left( \delta_{\alpha\beta}-\frac{Q_\alpha Q_\beta}{Q^2} \right)}\sum_{l,m}{e^{{\rm i}
                                  \mathbf{Q}\cdot(\mathbf{R}_l-\mathbf{R}_m)}}\times\\
                                  \sum_{M,M'}{\langle SM|\mathbf{S}_l^\alpha|S'M'\rangle}\langle S'M'|\mathbf{S}_m^\beta|SM\rangle\times\\
                                  \delta(\hbar\omega+E_S-E_{S'}),
\end{multline}
where $Z=\exp{\frac{3J}{4k_BT}}+3\exp{\frac{-J}{4k_BT}}$ is the partition function, $\alpha(\beta)=x, y, z$, and $\gamma$ the neutron magnetic moment, $r_0=e^2/m_ec^2$ the classical electron radius, $F(Q)$ the magnetic ion form factor, $\exp{(-2W)}$ the Debye-Waller factor, and $\mathbf{R}_l$ the position vector of the $l$th V ion in the molecule. $|SM\rangle=|\frac{1}{2}\frac{1}{2}SM\rangle$ and $|S'M'\rangle=|\frac{1}{2}\frac{1}{2}S'M'\rangle$ represent the state vectors of the initial and final electronic levels with energies $E_S$ and $E_{S'}$, respectively. The $\delta$ function may be replaced by a Gaussian line shape, due to line broadening, resulting from the relaxation effects and instrumental resolution.

By performing a powder average over the general cross section formula (10) in $\mathbf{Q}$ space, the $Q$ dependence of a magnetic transition is given by
\begin{equation}
I(Q)\propto |F(Q)|^2[1-f(QR)],
\end{equation}
where $f(x)=\sin{x}/x$ is a spherical Bessel function, and the interference term $1-f(QR)$ represents the separation $R$ of the two magnetic ions forming the dimer pair. Next we describe the main experimental results of Ref. \cite{D.A.Tennant}. The VODPO$_4\cdot\frac{1}{2}$D$_2$O has an orthorhombic crystal structure with space group $Pmmn$ at $T=10$ K, and with lattice parameters $a=7.4102(6)$ {\AA}, $b=9.5861(8)$ {\AA}, and $c=5.6873(7)$ {\AA} \cite{C.C.Torardi}. It is synthesized from V$_2$O$_5$, D$_3$PO$_4$, and D$_2$O, and the deuteration is necessary for performing the experiment, since one is interested in the magnetic INS and therefore the nonmagnetic scattering should be minimized by eliminating the H atoms. In addition, the deuteration has no measurable effect on the physical quantity. In the experiment the neutron scattering intensity is measured in the temperature range $10\leq T\leq200$ K as a function of energy transfer $E$ and wave vector transfer $Q$. By using the method of least-squares to fit a Gaussian profile with a linearly sloping background, the peak center is extracted at $J=7.81(4)$ meV, and the next-nearest-neighbor pathway (V-O-P-O-V shown in Figure 2) has a V-V separation $R=4.43(7)$ {\AA}. This value is close to 7.6 meV obtained from the measurement of magnetic susceptibility of VOHPO$_4\cdot\frac{1}{2}$H$_2$O \cite{J.W.Johnson}, which satisfies the Bleaney-Bowers formula well with the Land\'{e} factor $g=1.99$.
\begin{figure}[tbp]
\includegraphics[width=8.0cm]{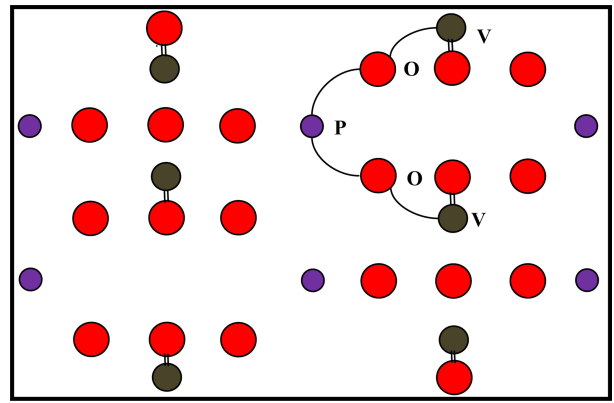}
\caption{(Color online) A schematic projection of VODPO$_4\cdot\frac{1}{2}$D$_2$O in the $ab$ plane. V ions are shown in grey, P in purple, and O in red. \label{Fig2}}
\end{figure}

Figure 3 displays the entanglement witness, concurrence, and quantum discord as a function of temperature in VODPO$_4\cdot\frac{1}{2}$D$_2$O. As expected, the correlations decrease with temperature. It is shown that at low temperature ($T\leq 10$ K), both concurrence and quantum discord remain maximal ($\thicksim$1) and are not affected by temperature, which is associated with the existence of a gap in the energy spectrum of the system. In the region $10\leq T\leq 53$ K, concurrence is always greater than the quantum discord, which implies that the quantum discord is not a sum of entanglement and some other nonclassical correlations. In the region $T>53$ K, quantum discord decays gradually with temperature, but is still
non-vanishing at room temperature. Therefore, it has a significant advantage over quantum entanglement, since the latter decreases quickly and
disappears at about $82.5$ K where the entanglement witness $|\langle\mathbf{S}_1\cdot\mathbf{S}_2\rangle|$ falls to 0.25. The critical temperature, where entanglement drops to zero, $T_c=\frac{J}{k_B\ln{3}}\approx 82.5$ K is much higher than 5 K in the copper nitrate CN[Cu(NO$_3$)$_2$2.5D$_2$O] \cite{G.Xu}. Similarly, at temperature below $T_c'=\frac{J}{k_B\ln{\frac{3+\sqrt{2}}{\sqrt{2}-1}}}\approx38.3$ K, the CHSH operator $|\langle\mathbf{B}\rangle|$ is higher than the local realistic limit, implying the quantum nonlocality. It is interesting to note that there is no monotonous relation between the Bell violation and quantum correlations. This means that an entangled state is not necessary a nonlocal state \cite{M.S.Kim}.
\begin{figure}[tbp]
\includegraphics[width=8.0cm]{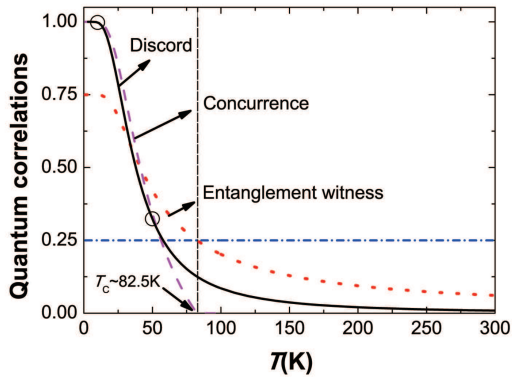}
\caption{(Color online) The temperature dependence of quantum correlations in VODPO$_4\cdot\frac{1}{2}$D$_2$O. The two open circles, marked for eye-catching, indicate the values of quantum discord derived at temperatures of 10 K and 50 K, respectively. \label{Fig3}}
\end{figure}

The spin-orbit interaction $\sum_{ij}{\mathbf{D}_{ij}\cdot(\mathbf{S}_i\times\mathbf{S}_j)}$ is crucial to the description of many antiferromagnetic systems such as Cu(C$_6$D$_5$COO)$_2\cdot$3D$_2$O, K$_2$V$_3$O$_8$, and Yb$_4$As$_3$, where $\mathbf{D}_{ij}$ is known as the Dzyaloshinski-Moriya vector \cite{I.Dzyaloshinsky,L.-C.Wang} and here we take $\mathbf{D}_{ij}=D\mathbf{e}_z$. By turning on the SO interaction, it is shown in Figure 4 that quantum entanglement and the critical temperature $T_c$ can greatly rise if the parameter $D$ is increased, i.e., larger entanglement can exist at moderately high temperature. This indicates that VODPO$_4\cdot\frac{1}{2}$D$_2$O is superior to many other candidates for quantum information process in solid-state systems, where quantum correlations are required at high temperature ($T\gg0$ K).

However, it is interesting to note that the SO interaction does not make a significant contribution to the quantum correlations for a family of Hamiltonians subject to the transformation \cite{L.A.Wu}
\begin{equation}
W=\prod_{j=1}{\exp{\left[ -{\rm i}(\theta_j S_j^x+\vartheta_j S_j^y+\phi_j S_j^z) \right]}},
\end{equation}
such that $ H'=WHW^\dag $, where $\theta_j$, $\vartheta_j$, and $\phi_j$ are independent angles, and the isotropic XY model with
\begin{equation}
W=\prod_{j=2,4,6,\ldots}{\exp{\left( {\rm i}\phi_j S_j^z \right)}}
\end{equation}
is a good physical example \cite{B.-Q.Liu}.

\begin{figure}[tbp]
\includegraphics[width=8.0cm]{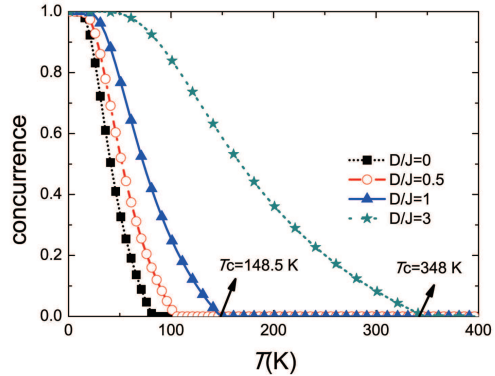}
\caption{(Color online) Quantum entanglement in VODPO$_4\cdot\frac{1}{2}$D$_2$O as a function of $T$ for different SO interactions, with $J=7.81$ meV. \label{Fig4}}
\end{figure}

\section{IV. Conclusions}

We analyze inelastic neutron scattering experiments and find the presence of macroscopic quantum correlations, quantified by entanglement and quantum discord, in the solid-state system VODPO$_4\cdot\frac{1}{2}$D$_2$O. Our study demonstrates that quantum correlations may play a broad generic role in the macroscopic world, and this low-dimensional magnetic material may be employed for quantum information processing, since that all quantum correlations remain approximately 1 at $T=10$ K, and the critical temperature for entanglement death is as high as $T_c=82.5$ K, in contrast to the materials with nuclear spins \cite{S.I.Doronin}, where entanglement can arise only at tenths of a microkelvin. On the other hand, quantum discord, another measure of non-classical correlations, can exist even at room temperature, enabling it a more effective resource for implementing quantum information processing. By introducing the spin-orbit interaction, we can even increase the critical temperature $T_c$ significantly. This feature is remarkable, implying that VODPO$_4\cdot\frac{1}{2}$D$_2$O may have practical implications for the study of fundamental issues of quantum mechanics and present promising applications in quantum technologies such as quantum computers.

\section*{ACKNOWLEDGMENTS}We would like to thank Marcelo S. Sarandy for illuminating discussions. This work was supported by the National Natural Science Foundation of China (Grants 11305150, 11105128, 91126001, 51231002, 11305156, 11305151). Lian-Ao Wu has been supported by the Basque Government (Grant IT472-10), and the Spanish MICINN (Projects No. FIS2012-36673-C03-01).


\begin{thebibliography}{99}

\bibitem{R.Horodecki} R. Horodecki, P. Horodecki, M. Horodecki, and K. Horodecki, Rev. Mod. Phys. {\bf 81}, 865 (2009).
\bibitem{S.M.Aldoshin} S. M. Aldoshin, E. B. Feldman, and M. A. Yurishchev, J. Exp. Theor. Phys. {\bf 107}, 804 (2008).
\bibitem{Z.-M.Wang} Z.-M. Wang, L.-A. Wu, C. Allen Bishop, Y.-J. Gu, M. S. Byrd, Phys. Rev. A {\bf 88}, 032303(2013).
\bibitem{T.Yu} T. Yu and J. H. Eberly, Phys. Rev. Lett. {\bf 93}, 140404 (2004).
\bibitem{H.Ollivier} H. Ollivier and W. H. Zurek, Phys. Rev. Lett. {\bf 88}, 017901 (2001).
\bibitem{T.Werlang} T. Werlang, C. Trippe, G. A. P. Ribeiro, and G. Rigolin, Phys. Rev. Lett. {\bf 105}, 095702 (2010).
\bibitem{P.Giorda} P. Giorda and M. G. A. Paris, Phys. Rev. Lett. {\bf 105}, 020503 (2010).
\bibitem{L.Mazzola} L. Mazzola, J. Piilo, and S. Maniscalco, Phys. Rev. Lett. {\bf 104}, 200401 (2010).
\bibitem{B.-Q.Liu} B.-Q. Liu, B. Shao, and J. Zou, Phys. Rev. A {\bf 82}, 062119 (2010); B.-Q. Liu, B. Shao, J.-G. Li, J. Zou, and L.-A. Wu, Phys. Rev. A {\bf 83}, 052112 (2011); B.-Q. Liu, B. Shao, and J. Zou, Commun. Theor. Phys. {\bf 56}, 46 (2011).
\bibitem{X.-X.Yi} L.-C. Wang, J. Shen, X.-X. Yi, Chin. Phys. B {\bf 20}, 050306 (2011).
\bibitem{X.-M.Lu} X.-M. Lu, Z.-J. Xi, Z. Sun, and X. Wang, Quantum Inf. Comput. {\bf 10}, 0994 (2010); J.-B. Yuan, L.-M. Kuang, and J.-Q. Liao, J. Phys. B {\bf 43}, 165503 (2010).
\bibitem{J.S.Bell} J.S.Bell, Physics {\bf 1}, 195 (1964).
\bibitem{A.Aspect} A. Aspect, P. Grangier, G. Roger, Phys. Rev. Lett. {\bf 49}, 91 (1982); A. Vaziri, G. Weihs, A. Zeilinger, Phys. Rev. Lett. {\bf 89}, 240401 (2002).
\bibitem{B.Terhal} B. Terhal, Phys. Lett. A {\bf 271}, 319 (2000).
\bibitem{M.Lewenstein} M. Lewenstein, B. Kraus, I. J. Cirac, P. Horodecki, Phys. Rev. A {\bf 62}, 052310 (2000).
\bibitem{F.A.Bovino} F. A. Bovino, G. Castagnoli, A. Ekert, P. Horodecki, C. M. Alves, and A. V. Sergienko, Phys. Rev. Lett. {\bf 95}, 240407 (2005).
\bibitem{P.Kok} P. Kok, W. Munro, K. Nemoto, T. Ralph, J. Dowling, and G. Miburn, Rev. Mod. Phys. {\bf 79}, 135 (2007).
\bibitem{G.Centi} G. Centi, Catal. Today (Special Issue) {\bf 16}, 1 (1993).
\bibitem{F.Trifiro} F. Trifiro, G. Centi, J. R. Ebner, V. M. Franchetti, Chem. Rev. {\bf 88}, 55 (1988).
\bibitem{E.Bordes} E. Bordes, Catal. Today {\bf 1}, 499 (1987).
\bibitem{D.A.Tennant} D. A. Tennant, S. E. Nagler, A. W. Garrett, T. Barnes, C. C. Torardi, Phys. Rev. Lett. {\bf 78}, 4998 (1997).
\bibitem{D.C.Johnston} D. C. Johnston, J. W. Johnson, J. Chem. Soc., Chem. Commun. {\bf 23}, 1720 (1985).
\bibitem{Y.Furukawa} Y. Furukawa, A. Iwai, A. Kumagai, K. Yakubovsky, J. Phys. Soc. Jpn. {\bf 65}, 2393 (1996).
\bibitem{J.Kikuchi} J. Kikuchi, T. Aoki, K. Motoya, T. Yamauchi, Y. Ueda, Physica B {\bf 284-288}, 1481 (2000).
\bibitem{J.W.Johnson} J. W. Johnson, D. C. Johnston, A. J. Jacobson, and J. F. Brody, J. Am. Chem. Soc. {\bf 106}, 8123 (1984).
\bibitem{M.Roca} M. Roca, P. Amor\'{o}s, J. Cano, M. D. Marcos, J. Alamo, A. Beltr\'{a}n-Porter, D.Beltr\'{a}n-Porter, Inorg. Chem. {\bf 37}, 3167 (1998).
\bibitem{H.J.Koo} H. J. Koo, M. H. Whangbo, P. D. VerNooy, C. C. Torardi, W. J. Marshall, Inorg. Chem. {\bf 41}, 4664 (2002).
\bibitem{S.Petit} S. Petit, S. Borshch, and V. Robert, J. Am. Chem. Soc. {\bf 124} 1744 (2002).
\bibitem{S.Petit2004} S. Petit, S. A. Borshch, and V. Robert, Inorg. Chem. {\bf 43}, 4210 (2004).
\bibitem{C.Brukner} \u{C}. Brukner, V. Vedral, and A. Zeilinger, Phys. Rev. A {\bf 73}, 012110 (2006).
\bibitem{G.Xu} G. Xu, C. Broholm, D. H. Reich, and M. A. Adams, Phys. Rev. Lett. {\bf 84}, 4465 (2000).
\bibitem{M.Yurishchev} M. Yurishchev, Phys. Rev. B {\bf 84}, 024418 (2011).
\bibitem{J.T.Haraldsen} J. T. Haraldsen, T. Barnes, J. L. Musfeldt, Phys. Rev. B {\bf 71}, 064403 (2005).
\bibitem{J.-M.Cai} J.-M. Cai, Z.-W. Zhou, G.-C. Guo, Phys. Lett. A {\bf 352}, 196 (2006).
\bibitem{X.Wang} X. Wang and P. Zanardi, Phys. Lett. A {\bf 301}, 1 (2002).
\bibitem{E.Lieb} E. Lieb, T. Schultz, and D. Mattis, Ann. Phys. (N.Y.) {\bf 16}, 407 (1961).
\bibitem{J.F.Clauser} J. F. Clauser, M. A. Horne, A. Shimony, and R. A. Holt, Phys. Rev. Lett. {\bf 23}, 880 (1969).
\bibitem{G.Jaeger} G. Jaeger, K. Ann, Phys. Lett. A {\bf 372} 2212-2216 (2008)
\bibitem{L.Henderson} L. Henderson and V. Vedral, J. Phys. A {\bf 34}, 6899 (2001); V. Vedral, Phys. Rev. Lett. {\bf 90}, 050401 (2003).
\bibitem{W.K.Wootters} W. K. Wootters, Phys. Rev. Lett. {\bf 80}, 2245 (1998).
\bibitem{G.Shirane} G. Shirane, S. M. Shapiro, and J. M. Tranquada, \emph{Neutron Scattering with a Triple-Axis Spectrometer Basic Techniques}, Cambridge University Press 2002.
\bibitem{H.U.Gudel} H. U. G$\ddot{u}$del, A. Stebler, and A. Furrer, Inorg. Chem. {\bf 18}, 1021 (1979).
\bibitem{C.C.Torardi} C. C. Torardi and J. C. Calabrese, Inorg. Chem. {\bf 23}, 1308 (1984); M. E. Leonowicz, J. W. Johnson, J. F. Brody, H. F. Shannon Jr., and J. M. Newsam, J. Solid State Chem. {\bf 56}, 370 (1985)

\bibitem{M.S.Kim} M. S. Kim, J. Y. Lee, D. Ahn, and P. L. Knight, Phys. Rev. A {\bf 65}, 040101(R) (2002).

\bibitem{I.Dzyaloshinsky} I. Dzyaloshinsky, J. Phys. Chem. Solids {\bf 4}, 241 (1958); T. Moriya, Phys. Rev. {\bf 117}, 635 (1960).
\bibitem{L.-C.Wang} L.-C. Wang, J.-Y. Y, X.-X. Yi, Chin. Phys. B {\bf 20}, 040305 (2011).
\bibitem{L.A.Wu} L.-A. Wu and D. Lidar, Phys. Rev. Lett. {\bf 91}, 097904 (2003).
\bibitem{S.I.Doronin} S. I. Doronin, A. N. Pyrkov, and \'{E}. B. Fel'dman, Pis'ma Zh. \'{E}ksp. Teor. Fiz. {\bf 85}, 627 (2007); Zh. \'{E}ksp. Teor. Fiz. {\bf 132}, 1091 (2007).

\end{thebibliography}
\end{document}